\documentclass[12pt]{article}
\usepackage{graphicx} % Required for inserting images
\usepackage{amsmath}
\usepackage[margin=1in]{geometry}
\usepackage{tikz}
\usepackage{setspace}
\usepackage{colortbl} % for color in tables
\usepackage{xcolor}   % for defining colors
\usepackage{authblk} 

\usepackage{hyperref}
\usepackage{verbatim, color,
amssymb, float, epsfig}
\usepackage{caption}
\usepackage{subcaption}
\usepackage[style=apa,doi=false,url=false,hyperref=true]{biblatex}
\usepackage{xr-hyper}
\usepackage{cleveref}

\newcommand{\ba}  {\begin{array}}
\newcommand{\ea}  {\end{array}}
\newcommand{\be}  {\begin{equation}}
\newcommand{\ee}  {\end{equation}}
\newcommand{\bea}  {\begin{eqnarray}}
\newcommand{\eea}  {\end{eqnarray}}
\newcommand{\nn}   {\nonumber} 
\newcommand{\ualpha}      {\mbox{\boldmath$\alpha$}}

\newcommand{\ugamma}      {\mbox{\boldmath$\gamma$}}

\newcommand{\utheta}      {\mbox{\boldmath$\theta$}}

\newcommand{\uiota}       {\mbox{\boldmath$\uiota$}}

\def\boxit#1{\vbox{\hrule\hbox{\vrule\kern6pt
 \vbox{\kern6pt#1\kern6pt}\kern6pt\vrule}\hrule}}

\providecommand{\keywords}[1]
{
  \small	
  \textbf{\textit{Keywords---}} #1
}

\makeatletter
\renewcommand*\env@matrix[1][\arraystretch]{%
  \edef\arraystretch{#1}%
  \hskip -\arraycolsep
  \let\@ifnextchar\new@ifnextchar
  \array{*\c@MaxMatrixCols c}}
\makeatother

\begin{document}            
\title{Robust Emax Model Fitting: Addressing Nonignorable Missing Binary Outcome in Dose-Response Analysis}

\author[1]{Jiangshan Zhang}
\author[2]{Vivek Pradhan}
\author[2]{Yuxi Zhao}

\affil[1]{Department of Statistics, University of California, Davis, CA 95616, USA}
\affil[ ]{jiszhang@ucdavis.edu}
\affil[2]{Global Biometrics and Data Management, Immunology and Inflammation, Pfizer Inc.,1 Portland Street, Cambridge, MA 02139, USA}
\affil[ ]{vivek.pradhan@pfizer.com}
\affil[ ]{yuxi.zhao@pfizer.com}
\date{}

\maketitle

\begin{abstract}
  The Binary Emax model is widely employed in dose-response analysis during drug development, where missing data often pose significant challenges. Addressing nonignorable missing binary responses—where the likelihood of missing data is related to unobserved outcomes—is particularly important, yet existing methods often lead to biased estimates. This issue is compounded when using the regulatory-recommended ‘imputing as treatment failure’ approach, known as non-responder imputation (NRI). Moreover, the problem of separation, where a predictor perfectly distinguishes between outcome classes, can further complicate likelihood maximization. In this paper, we introduce a penalized likelihood-based method that integrates a modified Expectation-Maximization (EM) algorithm in the spirit of Ibrahim and Lipsitz (\cite{ibrahim1996parameter}) to effectively manage both nonignorable missing data and separation issues. Our approach applies a noninformative Jeffreys’ prior to the likelihood, reducing bias in parameter estimation. Simulation studies demonstrate that our method outperforms existing methods, such as NRI, and the superiority is further supported by its application to data from a Phase II clinical trial. Additionally, we have developed an R package, \textit{ememax}, to facilitate the implementation of the proposed method.
\end{abstract}


\section{Introduction}
The dose-response relationship is a fundamental aspect of research in various applied statistics fields, particularly in clinical trials and bioinformatics. In real-world studies, incomplete data records are common due to reasons such as nonresponse to questionnaires, typographical errors, loss to follow-up, and data contamination. While missing data is often unavoidable, a common approach is to perform analysis using only complete records—a method known as complete case analysis (CC). This approach can be reasonable when the number of observations is large. However, in many scenarios, particularly in randomized controlled trials where sample sizes are often small, the application of complete case analysis can lead to biased estimates when analyzing dose-response relationships with binary outcomes (\cite{FIRTH1993,Maiti2008}).

The challenge of missing data has been a significant area of research for decades, with numerous statistical methods developed to address it. Various missing data mechanisms, such as missing completely at random (MCAR) and missing at random (MAR), have been extensively studied, and corresponding methodologies have been proposed. It is well known that if the missing data mechanism is not appropriately modeled, parameter estimates may be biased (\cite{little2019statistical}). While much of the research has focused on missing covariates, less attention has been given to scenarios where some response values are missing. When the likelihood of missingness depends on the unobserved values of the response, this is referred to as nonignorable missing data and 
 existing literature addressed handling it in the context of survey research (\cite{Little1982}). In this article, we explore the implications of nonignorable missing data within the context of a dose-response model with binary outcomes. 

The sigmoid Emax model is commonly used in clinical trials to explore the binary dose-response relationship. Let $n$ denote the total sample size, and suppose $y_i$ denotes the binary outcome for the $i$-th patient randomized to a dose $Dose_i$, where $Dose_i$ is one of the predefined dose levels from a set of $J$ levels $\{D_1,\cdots,D_J\}$, for $i=1,\cdots,n$. Without loss of generality, let $P(y_i =1|Dose_{i})= \pi_{i}$ be the probability of success for patient $i$ after dosage. Under this setup, the sigmoid four-parameter Emax model can be written as the following 
\be
log\left(\frac{\pi_{i}}{1-\pi_{i}}\right)=E_0+\frac{E_{max}\times Dose_{i}^{\lambda} }{ED_{50}^{\lambda} +Dose_{i}^{\lambda} }
\ee  
where $E_0$ is the expected logit of dose effect at $Dose_{i}=0$, with $Dose_{i}=0$ often being considered as placebo; $E_{max}$ is the expected logit of the maximum achievable effect (at infinite dose); $ED_{50}$ is the dose that produces half-maximal effect $E_{max}/2$; and $\lambda$ is the slope factor or Hill parameter, determining the steepness of the dose-response curve. Even though the four-parameter Emax model is available in the literature, according to meta-analysis studies by Thomas et al. (\cite{Thomas20172}), oftentimes in practice the three-parameter Emax model fits well on most real-world data, where the Hill parameter is assumed to be $1$ (\cite{Kirby2011,Wu2017}). Hence by putting $\lambda=1$ the Emax model is reduced to

\be
\log\left(\frac{\pi_{i}}{1-\pi_{i}}\right)=E_0+\frac{E_{max}\times Dose_{i} }{ED_{50}+Dose_{i} }.
\label{eq:3param}
\ee
Define $\utheta=( E_0, ED_{50},E_{max})^\top$, then from (\ref{eq:3param}), the success probability $\pi_{i}$ for patient $i$ can be written as
\bea
P(y_{i} = 1|\utheta, Dose_{i})&= &f(y_{i}|\utheta, Dose_{i}) \nn \\
&=&\exp\left(E_0+\frac{E_{max}\times Dose_{i} }{ED_{50}+Dose_{i} }\right)/\left(1+ \exp(E_0+\frac{E_{max}\times Dose_{i} }{ED_{50}+Dose_{i} })\right).
\eea
Given a dataset, finding the estimate $\widehat{\utheta}$ of the parameter $\utheta$ is the main point of interest in clinical trials for exploring dose-response characteristics. Once $\widehat{\utheta}$ is obtained, one can easily estimate the success rates of different doses and draw statistical inferences of the corresponding dose populations. The estimate $\widehat{\utheta}$ of the parameter $\utheta$ can be obtained by maximizing the likelihood (actually log-likelihood) given in the following:

\be
L(\utheta, Dose)= \prod_{i=1}^{n} f(y_i|\utheta, Dose_{i}).
\ee 

In clinical trials, it is common for binary outcomes to be missing for some patients receiving either an experimental or placebo dose. Given the typically small sample sizes in these studies, complete case analysis is often impractical. A standard practice, endorsed by the FDA and other regulatory agencies, is to impute all missing values as "treatment failures." This method, known as non-responder imputation (NRI) (\cite{ONeill2012}), is widely used but has well-documented limitations. NRI can introduce substantial bias into estimates and inferences, particularly when the missing data mechanism is nonignorable (\cite{nri123}).

Another commonly used method for addressing missing data is multiple imputation (MI). This approach involves a two-stage process: first, generating several plausible imputed datasets based on assumed data distributions, and second, combining the results using a predefined pooling rule, as developed by Rubin (\cite{Rubin1987}). However, MI typically assumes that the missingness mechanism does not depend on the unobserved data, an assumption that is violated under nonignorable missingness (\cite{GiustiCaterina2011AAoN}). Consequently, MI can also result in biased estimates and inferences. Some modified MI methods have been proposed to address missing not at random (MNAR) scenarios by model selection approach (\cite{Galimard2016, Galimard2018}) or pattern-mixture model approach (\cite{vanBuuren1999}), but these methods often rely on assumptions of imputation model, especially the difference between distributions of respondent and non-respondent. This limits their flexibility in practical applications, where variables associated with missingness can vary widely. Moreover, the performance of MI heavily depends on the chosen pooling methods after generating and analyzing imputed datasets, which can be challenging to determine in practice (\cite{Carpenter2007}). Therefore, developing new approaches to fit the Emax model with incomplete data, particularly when the missing data mechanism is nonignorable, remains a critical issue in medical research.

In small or medium sample size settings for fitting binary outcome models, a common issue that can arise is the nonconvergence of estimates, a phenomenon known as “separation” (\cite{ALBERT1984}). Complete separation occurs when a single covariate or a linear combination of covariates perfectly predicts the outcome, leading to divergence in the estimation process. Even if complete separation does not occur, the presence of quasi-complete separation—where a subset of subjects’ responses is perfectly predicted—can still cause estimation challenges (\cite{altman2004numerical}). These situations, though not uncommon in biomedical datasets, are often overlooked. For example, current methods and packages that use likelihood estimation for dose-response or Emax models, such as the \textit{ClinDR} package in R developed by Thomas (\cite{Thomas2017}) and the \textit{Dosefinding} package by Bornkamp et al. (\cite{Bornkamp2010}), do not account for the issue of separation.

In the framework of generalized linear models, Heinze and Schemper (\cite{heinze2002solution}) and Heinze (\cite{heinze2006comparative}) demonstrated that Firth’s method (\cite{FIRTH1993}), initially designed to reduce the bias of maximum likelihood estimates (MLE), can also effectively address the problem of separation. However, their work did not account for the presence of missing data. Extending this approach, Maity and Pradhan (\cite{maiti2009bias,Maity2018}) applied Firth’s method to reduce bias and adjust for separation under a nonignorable missing data mechanism within the logistic regression framework. Despite these advances, their work did not address nonlinear covariate relationships, such as those encountered in the Emax model.

In the field of dose-response relationships, no prior research has addressed the issue of nonignorable missing data within the Emax model framework. Additionally, there is no existing literature on bias reduction and separation in the Emax model, even with complete data. In this article, we introduce a novel approach that integrates Firth-type adjustments with the Emax model to handle nonignorable incomplete binary data, effectively addressing both nonignorable missingness and potential separation. The remainder of the paper is organized as follows: Section 2 outlines the proposed approach and details the derivation of the estimation algorithm. Section 3 explores the simulation settings and presents the results. Section 4 discusses a real-world application, providing insights into the practical implementation of the proposed method. Finally, Section 5 offers a discussion of the findings and suggests potential directions for future research. Additionally, we have developed an R package, \textit{ememax}, that implements our methods.

\section{Method}
\subsection{Weighted EM Procedure of Ibrahim and Lipsitz (IL)}

Let $ \mathbf{r} $ be the missing indicator vector whose $i$-th element is defined as
	\[
	r_i = \left\{ 
	\begin{array}{l l}
	1 &\quad \text{if $y_i$ is missing}\\
	0 &\quad \text{if $y_i$ is observed}
	\end{array} \right.\]
	and is generated by 
	\[
	P(r_i = 1| z_i,\ualpha) = p_i=\frac{\exp\left(z_i^\top\ualpha\right)}{1 + \exp\left(z_i^\top\ualpha\right)}, \quad i = 1, 2, \ldots, n
	\]
where $ z_i = (x_i^\top,Dose_{i}, y_i)^\top $ is the covariate vector with $x_i$ to be a $(p+1)\times 1$ covariate vector of interest including an intercept term, and $ \ualpha = (\alpha_0,...,\alpha_p, \alpha_{(p+1)}, \alpha_{(p+2)})^\top $ is the corresponding parameter vector. If $ \alpha_{p+2} = 0 $, the missing data mechanism does not depend on $y_i$, and hence the missing mechanism is ignorable. However, if $ \alpha_{p+2} \neq 0 $, the missing mechanism depends on $y_i$ and is therefore nonignorable. Note that when $ \ualpha $ is a null vector, the missing mechanism is MCAR. For fitting binary regression model with nonignorable missing values, Ibrahim and Lipsitz (\cite{ibrahim1996parameter}) proposed an EM algorithm to compute the estimate of the regression coefficient. Following Ibrahim and Lipsitz (\cite{ibrahim1996parameter}) combined with the binary Emax model, the joint likelihood can be written as,
\be
f(r,y|\ualpha,\utheta, Dose,x)= \prod_{i=1}^{n} f(y_i|\utheta, Dose_{i}) f(r_i|z_i, \ualpha)
\label{eq:lkhd}
\ee
The maximum likelihood of $(\ualpha,\utheta )$ of (\ref{eq:lkhd}) can be obtained via the EM algorithm by maximizing the expected log-likelihood. The E-step of the $i$-th individual's contribution can be written as 
\begin{align}
	E[l(\utheta, \ualpha | z_i, r_i)] = \left\{
	\begin{array}{l l}
	\sum_{y_i = 0}^{1} l(\utheta, \ualpha | z_i, r_i) f(y_i | r_i, Dose_{i}, x_i, \utheta, \ualpha) & \quad \text{if $y_i$ is missing} \\
	l(\utheta, \ualpha | z_i, r_i) & \quad \text{if $y_i$ is observed}\\
	\end{array}
	\right .
	\end{align}
where $ f(y_i | r_i, Dose_{i}, \utheta, \ualpha)$ is the conditional distribution of the missing outcome given the observed data, and $l(\utheta, \ualpha | z_i, r_i)$ is the log-likelihood of the $i$-th individual given in (\ref{eq:lkhd}). In the above expectation, let $w_{i y_i}=f(y_i | r_i, Dose_{i},x_i, \utheta, \ualpha)$, which can be considered as weight. This can be written further using Bayes theorem as 
\bea
f(y_i | r_i, Dose_{i}, x_i, \utheta, \ualpha) = w_{i y_i} &=& \frac{ f(r_i,y_i|\ualpha,\utheta, x_i, Dose_{i})}{ \sum_{y_i=0}^{1}f(r_i,y_i|\ualpha,\utheta,x_i, Dose_{i})} \nn \\
&=&\frac{ f(y_i|\utheta, Dose_{i}) f(r_i|z_i, \ualpha)}{\sum_{y_i=0}^{1} f(y_i|\utheta, Dose_{i}) f(r_i|z_i, \ualpha)}
\eea
Therefore, for all data with $i=1,\cdots,n$, the $(t+1)$-th iteration of the E-step can be expressed as	
\begin{equation} 
 \begin{split}
     Q(\utheta, \ualpha|\utheta^{(t)}, \ualpha^{(t)}) & = \sum_{i=1}^{n} \sum_{y_i = 0}^1 w_{i y_i}^{(t)} l(\utheta, \ualpha | z_i, r_i,Dose_{i},y_i)\\
     & = \sum_{i=1}^{n} \sum_{y_i = 0}^1 w_{i y_i}^{(t)} \left\{ l(\utheta |Dose_{i},y_i) 
 + l(\ualpha | z_i, r_i)\right\}
 \end{split}
 \label{M_step}
	\end{equation}
where the $t$-th stage weights of E-step are defined as,
	\begin{align}
	w_{i y_i}^{(t)} &= \left\{
	\begin{array}{l l}
	\frac{f(y_i | Dose_{i}, \utheta^{(t)}) f(r_i | z_i, \ualpha^{(t)})}{\sum_{y_i = 0}^{1} f(y_i | Dose_{i}, \utheta^{(t)}) f(r_i | z_i, \ualpha^{(t)})} & \quad \text{if $y_i$ is missing} \\
	1 & \quad \text{if $y_i$ is observed}\\
	\end{array}
	\right.
    \label{weight}
	\end{align}
Since (\ref{M_step}) is the sum of two equations involving parameter $\ualpha$ for the logistic model and parameter $\utheta$ for the Emax model, and both parameters $\ualpha$ and $\utheta$ are independent by assumption, for M-step, we can maximize these separately. The maximization of $l(\ualpha | z_i, r_i)$ is straightforward, which can be done iteratively by the Newton-Raphson method with the following updating equation:
\begin{equation}
    \ualpha^{(s+1,t)}=\ualpha^{(s,t)} + I(\ualpha^{(s,t)})^{-1}U(\ualpha^{(s,t)})
    \label{eq::updating}
\end{equation}
where $\ualpha^{(s,t)}$ is the estimate of $\ualpha$ at $s$-th iteration of the Newton-Raphson procedure within $t$-th iteration of the EM algorithm, $I(\ualpha)=Z^{T}V(\ualpha)Z$ is the information matrix at $\ualpha$ with $V(\ualpha)=\mathrm{diag}(w_{i y_i} p_i(1-p_i))$, and $U(\ualpha)$ is the score function defined as in the following:
\begin{equation}
    U(\ualpha) = \sum_{i=1}^n\sum_{y_i=0}^1 w_{i y_i} z_i(r_i-p_i).
\end{equation}
The entire procedure can be implemented using standard software packages that fit logistic regression with specified weight options. Next, for the M-step of $l(\utheta |Dose_{i},y_i)$, consider the Newton-Raphson method to find the maximizer. After some algebra, the score function reduces to
\begin{equation}
    U(\utheta) = \sum_{i=1}^n\sum_{y_i=0}^1 w_{i y_i} (y_i-\pi_i) \nabla \eta(Dose_{i},\utheta)
\end{equation}
where $\eta(Dose_{i},\utheta) = E_0+E_{max}\times Dose_{i}/({ED_{50}+Dose_{i} })$, $\nabla$ is the differential operator with respect to $\utheta$, and
$$\nabla \eta(Dose_{i},\utheta)=\left(1, \frac{Dose_{i}}{Dose_{i}+ED_{50}}, -\frac{Dose_{i}\times E_{max}}{(Dose_{i}+ED_{50})^2} \right)^{T}.$$
The Hessian matrix is:

\begin{equation}
    H(\utheta) =\sum_{i=1}^n\sum_{y_i=0}^1 w_{i y_i} \left ((\pi_i-1)\pi_i  \nabla \eta(Dose_{i},\utheta)^\top\nabla \eta(Dose_{i},\utheta) - A_i(\utheta) \right )
\end{equation}
where
$$A_i(\utheta)=\begin{pmatrix}[1.5]
    0 & 0 & 0\\
    0 & 0 & \frac{(y_i-\pi_i)Dose_{i}}{(ED_{50}+Dose_{i})^2}\\
    0 & \frac{(y_i-\pi_i)Dose_{i}}{(ED_{50}+Dose_{i})^2} & - \frac{2(y_i-\pi_i)Dose_{i}\times E_{max}}{(ED_{50}+Dose_{i})^3}
\end{pmatrix}.$$
The observed information matrix can be obtained by taking the negative value of the Hessian matrix, and the updating equation of $\utheta$ is similar as \eqref{eq::updating} of $\ualpha$. It is pertinent to point out that $A_i(\utheta)$ is the additional term due to non-linear model setting compared to the Hessian matrix of logistic regression of $i$-th observation when weight $ w_{i y_i}=1$. Finally, by repeating the E-step and M-step until convergence, we can get the estimates of $\utheta$ and $\ualpha$. 
%Note that in each E-step, the weights are updated using \eqref{weight}.

\subsection{Firth-type Bias Reduction on Ibrahim and Lipsitz (FIL)}
Firth (\cite{FIRTH1993}) introduced a modification to the score function of the likelihood to reduce the bias of the maximum likelihood estimator (MLE) in small sample settings. Firth demonstrated that when the parameter in question is the canonical parameter of a full exponential family—such as in the logistic regression model for the missing indicator model with $r$ —the modification of the score function is equivalent to applying a Jeffreys’ invariant prior to the likelihood function. In the spirit of Firth, to obtain explicit Firth-type bias-reduced estimates of the parameter $\utheta$ for the Emax model with missing outcome, the joint likelihood corresponding to (\ref{eq:lkhd}) can be modified as follows:

\[
f^{*}(r,y \mid \ualpha, \utheta, x, Dose) = \prod_{i=1}^{n} f^{*}(y_i \mid \utheta, Dose_{i}) f^{*}(r_i \mid z_i, \ualpha),
\label{eq:lkhdpenalized}
\]
where the Firth-type bias reduction is achieved by penalizing each likelihood component on the right-hand side by multiplying it with the Jeffreys’ invariant prior as the penalty term. The penalized log-likelihood function for modeling nonignorable missingness is:

\begin{equation}
l^*(\ualpha \mid r,z) = l(\ualpha \mid r,z) + \frac{1}{2}|I(\ualpha)|,
\label{pll:r}
\end{equation}
where $|I(\ualpha)|$ is the determinant of the observed information matrix. The penalized log-likelihood for the Emax model part is given by:

\begin{equation}
l^*(\utheta \mid y, Dose) = l(\utheta \mid y, Dose) + \frac{1}{2}|I(\utheta)|,
\label{pll:y}
\end{equation}
and the joint penalized log-likelihood can be obtained by combining \eqref{pll:r} and \eqref{pll:y}, utilizing the independence of \( \utheta \) and \( \ualpha \). The bias-reduced MLE can then be obtained by maximizing this joint penalized log-likelihood.

Subsequently, we apply IL method with penalized joint log-likelihood, and the E-step of the EM becomes
\begin{equation}
    Q^*(\utheta,\ualpha|\utheta^{(t)},\ualpha^{(t)}) = \sum_{i=1}^n\sum_{y_i=0}^1 w_{i y_i}^{(t)} \left \{l^*(\utheta|y_i,Dose_{i}) + l^*(\ualpha|r_i,z_i)) \right \}.
\end{equation}
The modification in the E-step as shown above leads to a change in the M-step to obtain the maximizer as well. For $ l^*(\ualpha|r_i,z_i)$, the modified score function $U^*(\ualpha)$ takes the form as
\begin{equation}
    U^*(\ualpha) = \sum_{i=1}^n\sum_{y_i=0}^1 w_{i y_i} z_i [y_i-p_i+h_i(1/2-p_i)],
\end{equation}
where $h_i$ is the $i$-th diagonal elements of the hat matrix $V^{1/2}Z(Z^\top VZ)^{-1}Z^\top V^{1/2}$. For $l^*(\utheta|y_i,Dose_{i})$, the corresponding score function $U^*(\utheta)$ is formulated as:

\begin{equation}
    U^*(\utheta) = \sum_{i=1}^n\sum_{y_i=0}^1 w_{i y_i} (y_i-\pi_i) \nabla \eta(Dose_{i},\utheta) + \mathrm{tr}\left(I(\utheta)^{-1} \frac{\partial I}{\partial \utheta} \right ).
    \label{score::y}
\end{equation}

Finally, the estimator can be obtained via any iterative procedures such as Newton-Raphson, Gauss-Newton, and Fisher scoring.

\subsection{Variance Estimator and Confidence Interval}
Let $\ugamma=(\utheta,\ualpha)$ denote the parameter for estimation via EM, and $\hat{\ugamma}$ denote the final estimator obtained from EM. The approximate variance-covariance matrix of $\hat{\ugamma}$ for both IL and FIL methods can be estimated via observed information matrix $I(\ugamma)$. In the EM setting, $I(\ugamma)$ can be estimated using Louis (\cite{Louis1982}) as below:
\begin{equation*}
    \begin{split}
     I(\ugamma) = &-H(\ugamma) - \mathrm{E}\left[S(\ugamma|y,r,Dose,z)S(\ugamma|y,r,Dose,z)^\top\right] \\
     &+\mathrm{E}\left[S(\ugamma|y,r,Dose,z) \right]\mathrm{E}\left[S(\ugamma|y,r,Dose,z) \right]^\top 
\end{split}
\end{equation*}
where $S(\ugamma|y,r,Dose,z)$ is the complete-data score vector, and all the expectations are with respect to the conditional distribution of missing outcome given the observed data. Incorporating with the IL or FIL setting, we define
\begin{align*}
    S_i(\ugamma) &= \frac{\partial}{\partial \ugamma} l(\ugamma|y,r,Dose,z)\\
    \dot q_i(\ugamma) &= \sum_{y_i=0}^1 w_{iy_i} S_i(\ugamma).
\end{align*}
Thus, the estimated observed information matrix of $\ugamma$ giving observed data is
\begin{equation}
    \begin{split}
       I(\hat{\ugamma}) = &-H(\hat{\ugamma}) - \sum_{i=1}^n\sum_{y_i=0}^1 \hat{w}_{i y_i} S_i(\hat{\ugamma}) S_i(\hat{\ugamma})^\top + \sum_{i=1}^n \dot q_i(\hat{\ugamma})\dot q_i(\hat{\ugamma})^\top
    \end{split}
    \label{eq:vcov}
\end{equation}
where $\hat{w}_{i y_i}$ is the estimate at convergence of EM. Note that all the quantities in \eqref{eq:vcov} can be obtained easily from the M-step as byproducts. We can get a consistent estimator of variance-covariance matrix as the inverse of $I(\hat{\ugamma})$.

With the variance-covariance matrix estimator $I(\hat{\ugamma})^{-1}$, confidence intervals for the $\ugamma$ can be constructed. The standard errors of parameters can be obtained by taking the square root of the diagonal elements of the variance-covariance matrix, and the 100(1-$\alpha$)\% confidence interval based on the normal approximation for parameter $\gamma_i$ is:
\begin{equation}
    (\hat{\gamma}_i -z_{\alpha/2}\hat{s}_{\gamma_i},\hat{\gamma}_i +z_{\alpha/2}\hat{s}_{\gamma_i})
\end{equation}
where $\hat{\gamma}_i$ is the estimator of ${\gamma}_i$, $z_{\alpha/2}$ is the $1-\alpha/2$ quantile of the standard normal distribution, and $\hat{s}_{\gamma_i}$ is the estimated standard error of $\hat{\gamma}_i$.

\section{Simulation Study}
To evaluate the performance of the proposed methodology, we conducted a series of simulation studies based on a general Phase-II dose-response clinical trial setting. We investigated the effect of sample size on the estimation process by considering sample sizes of n = 150, 250, 350, and 450. The total sample size was evenly distributed across five different treatment dose arms: Dose = (0, 7.5, 22.5, 75, 225), ensuring equal sample sizes in each treatment arm. The success rate for response in the placebo arm $(Dose_1 = 0)$ was set at 10\%, the maximum success rate with an infinite dose at 80\%, and the dose achieving a half-maximal effect at 7.5. This corresponds to $E_0 = \text{logit}(0.1)$, $E_{max} = \text{logit}(0.8) - \text{logit}(0.1)$, and $ED_{50} = 7.5$.

The response variable $y_i$ was generated from a Bernoulli distribution with success probability $\pi_i$ as defined in \eqref{eq:3param}, for $i = 1, \dots, n$. After generating $y_i$, the nonignorable missing data indicator $r_i$ was generated using a Bernoulli distribution with a missing probability $p_i$, where $p_i = {\exp(z_i^\top \ualpha)}/({1 + \exp(z_i^\top \ualpha))}$. The covariate vector was defined as $z_i = (1, x_{i1}, x_{i2}, Dose_{i}, y_i)^\top$, and $\ualpha = (\alpha_0, \alpha_1, \alpha_2, \alpha_3, \alpha_4)^\top$. We sampled $x_{i1}$ and $x_{i2}$ from two independent standard normal distributions. If $r_i$ was generated as 1, the corresponding $y_i$ value was masked as ‘NA’, indicating a missing response. The nonignorable nature of the missing mechanism was ensured by setting $\alpha_4 \neq 0$. The overall rate of missing responses was controlled by selecting appropriate values for $\ualpha$. For instance, to achieve an overall missing rate of approximately 15\%, we set $\ualpha = (-2.5, 3, 0, -0.05, 1)^\top$. A noteworthy aspect is that $\alpha_2$ was set to 0 to introduce a degree of model mis-specification when using the full $z_i$ vector to predict missingness.

Table \ref{Table::diff_N} summarizes the simulation results comparing five different estimation methods: Complete Case (CC), Non-Responder Imputation (NRI), Multiple Imputation (MI), Ibrahim-Lipsitz (IL), and Firth-Adjusted IL (FIL), based on $N = 1000$ replications. For each of the simulated data with missing values, MI was performed with $m = 100$ imputations using the \textit{mice} package, employing predictive mean matching and lasso-logistic regression imputation methods along with the Namard-Rubin pooling rule. The CC and NRI analyses were conducted using the \textit{ClinDR} package. For point estimation, we report the average estimated value of $\theta_i$, $\hat{\theta_i} = \frac{1}{s}\sum \hat{\theta_i}^{(k)}$ for $i = 1, 2, 3$ corresponding to the three-parameter Emax model, where $\hat{\theta_i}^{(k)}$ is the estimate of $\theta_i$ in the $k$-th replication, and $s$ is the number of valid estimates out of $N$ replications. Additionally, we report the mean bias error (MBE) as $\mathrm{MBE} = \frac{1}{s}\sum (\hat{\theta_i}^{(k)} - \theta_i)$, the mean squared error (MSE) as $\mathrm{MSE} = \frac{1}{s}\sum (\hat{\theta_i}^{(k)} - \theta_i)^2$, and the mean estimated standard error for $\hat{\theta}_i$, $\hat{s}_{\theta_i} = \frac{1}{s}\sum \hat{s}_{\theta_i}^{(k)}$. For confidence interval estimation, we report the coverage probability CP at 95\% confidence level using $\text{CP} = \frac{1}{s}\sum I(\hat{\theta_i}^{(k)})$, where $I(\cdot)$ is the indicator function for whether $\theta_i$ falls within the estimated confidence interval. We also report the mean estimated interval length (Est. length) as $ \text{Est. length} = \frac{1}{s}\sum \lambda(\hat{\theta_i}^{(k)})$, where $\lambda(\hat{\theta_i}^{(k)})$ represents the length of the confidence interval for $\hat{\theta_i}^{(k)}$.

\textbf{Table \ref{Table::diff_N} is approximately here.}

As shown in Table \ref{Table::diff_N}, NRI estimators exhibit severe MBE and MSE compared to other methods across all scenarios, leading to lower CP despite having lower mean estimated standard errors. For CC estimators, although the MSE occasionally performs better than IL, and they are nearly unbiased for $ED_{50}$, the MBEs for $E_{max}$ and $E_0$ are consistently large, resulting in a highly biased estimator. MI estimators show some reduction in bias for point estimates across all scenarios compared to CC and NRI; however, the bias remains substantial. Furthermore, the standard error for MI is underestimated, leading to a reduced nominal coverage for the confidence intervals. FIL consistently outperforms IL, especially in estimating $E_0$ and $E_{max}$ in terms of MBE. Concerning MSE and mean estimated standard errors, FIL estimators consistently achieve the lowest values across all scenarios. Due to the reduced standard errors, FIL also produces the narrowest confidence intervals. If we consider the square root of MSE as the true simulated standard error of the parameters, the estimated standard errors for both IL and FIL tend to be slightly overestimated, with this overestimation being more pronounced in smaller sample sizes. This leads to conservative confidence intervals and an overestimation of nominal coverage.

The FIL method for bias correction is particularly valuable when the sample size is small and separation is encountered in the missingness pattern. Additional simulations were performed where missingness was more correlated with the dose, leading to severe separation in the placebo treatment arm (results reported in \cref{sup-Table::S1}). In this scenario, all methods perform worse due to the systematic bias introduced by the loss of extreme scorers (success cases) in the placebo group. However, FIL and IL still outperform CC and NRI, with FIL consistently achieving better results. Notably, IL significantly underestimates the nominal coverage, whereas FIL maintains a reasonable coverage probability. Additionally, the standard errors estimated by FIL are consistently smaller than those from IL, demonstrating the effectiveness of the Firth-type method in mitigating the effects of separation.

\begin{figure}[htbp]
    \centering
    \includegraphics[width=1\textwidth]{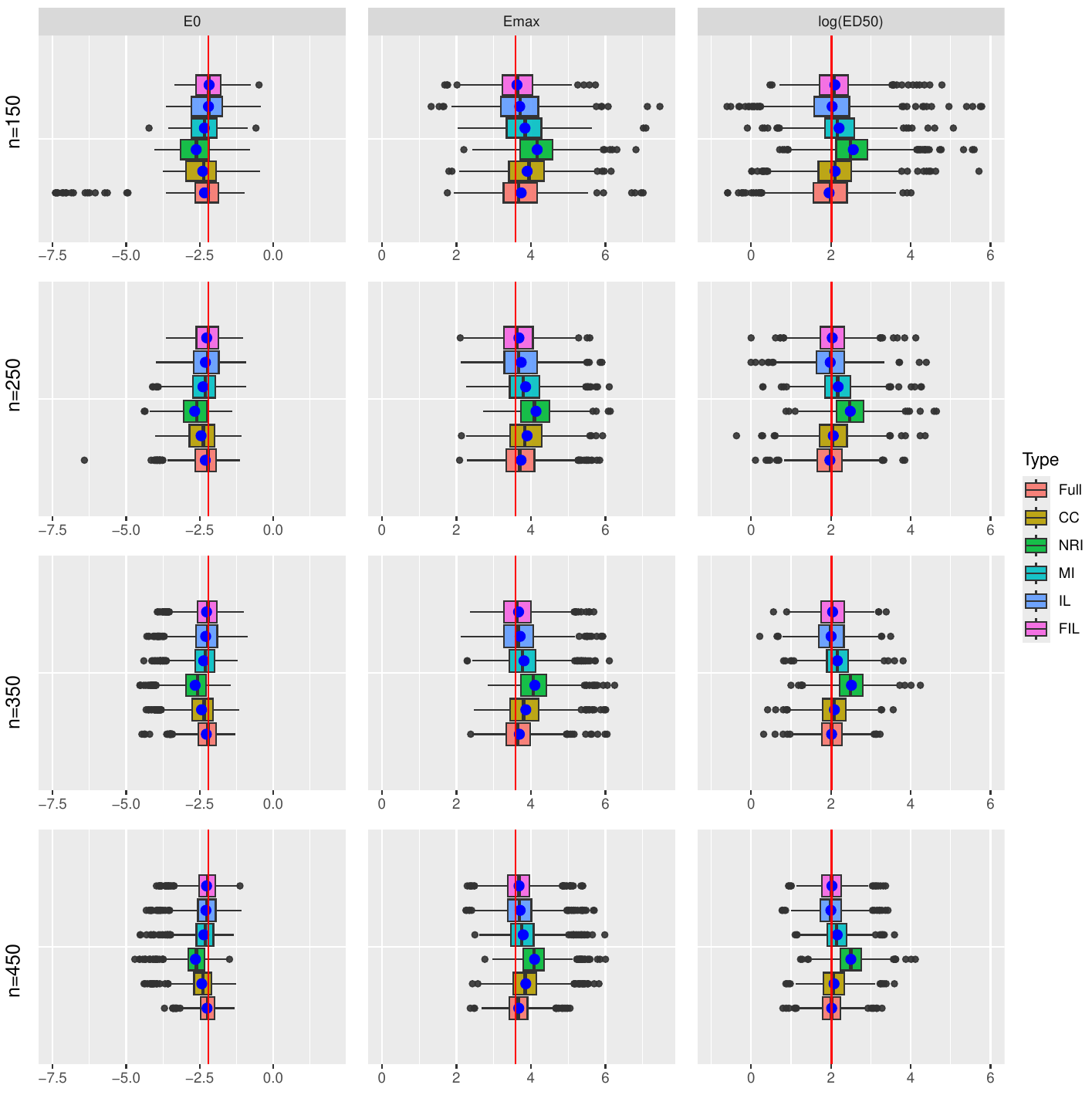}
    \caption{Boxplots comparing the distribution of point estimates with true parameter $E_0=-2.197$, $E_{max}=3.584$, and $ED_{50}=2.015$, based on 1000 replications and missing rate approximately 15\%.}
    \label{fig:mis15_comp}
\end{figure}

\textbf{Figure \ref{fig:mis15_comp} is approximately here.}

Figure \ref{fig:mis15_comp} illustrates the distribution of point estimates for different methods, along with estimates based on the full data without missingness, across varying sample sizes. The true value of $\utheta$ is indicated by a vertical line on the boxplot, and the mean estimates are shown as blue dots. The plot highlights that FIL effectively reduces bias due to small sample sizes and separation, particularly when $n = 150$. Even with the full dataset, separation due to small sample size results in unstable MLE estimates, characterized by large variations and outliers. However, the application of Jeffreys' prior modification in FIL, which introduces strong convexity in the likelihood function with respect to the parameter of interest, mitigates the constancy of the likelihood due to separation, leading to more reliable estimates. Additionally, while both IL and FIL perform similarly in terms of median estimates, the box lengths (indicating variability) for FIL are consistently narrower than those for IL.

Further simulations were conducted with varying missing rates and a fixed sample size of $n = 350$. The missing rates were set at 10\%, 15\%, 25\%, and 30\%, with different $\ualpha$ combinations used to produce the varying rates. As shown in Table \cref{sup-Table::S2}, the results are consistent with those in \ref{Table::diff_N}. FIL outperforms IL, and both methods provide better estimates than CC, NRI, and MI. Additionally, FIL consistently achieves the lowest mean estimated standard errors, with coverage probabilities close to the desired 95\% nominal level.

\section{Real Data Analysis}
This section presents an example of fitting a dose-response model using data from the TURANDOT study (\cite{TURANDOT}), a Phase II randomized, double-blind, placebo-controlled clinical trial for ulcerative colitis in patients with moderate to severe disease. In this study, 357 patients were randomly assigned to either a placebo group or one of four active dose groups: 7.5 mg, 22.5 mg, 75 mg, and 225 mg. As reported by Vermeire et al. (\cite{TURANDOT}), a non-monotone dose-response profile was observed, with lower efficacy in the highest dose group (225 mg). Among the 357 patients, 73 received placebo, 71 received 7.5 mg, 72 received 22.5 mg, 71 received 75 mg, and 70 received 225 mg. The primary endpoint was clinical remission at Week 12, which included several missing values believed to be nonignorable, as summarized in Table \ref{Table::turandot_descriptive}.

\textbf{Table \ref{Table::turandot_descriptive} is approximately here.}

We fitted the Emax model, assuming all missing values in remission were nonignorable. For modeling the missingness indicator, we performed model selection based on the Akaike Information Criterion (AIC) with a lot of covariates and their interactions, and the final model included the following covariates: remission response (y), dose, Mayo score at baseline (MCSBASE), C Reactive Protein score at baseline (CRPBASE), age, sex, immune suppressant status (IS), steroid use history (SD), and Acetylsalicylic acid use history (ASA). We compared our proposed method to complete case (CC) analysis, non-responder imputation (NRI), and multiple imputation (MI). Note that for MI, we used the imputation model which is the same as that used in IL and FIL excluding the outcome variable. The results of the dose-response relationship using different methods are shown in Table \ref{Table::turandot_analysis}.

\textbf{Table \ref{Table::turandot_analysis} is approximately here.}

We observe that the IL method yields an unstable estimate for log($ED_{50}$), with a very wide 95\% confidence interval. This instability is likely due to separation issues in the logistic regression for the missingness indicator. Similarly, the MI method also produces a high standard error for estimating log($ED_{50}$). In contrast, the FIL method provides a more stable estimate due to the use of penalized maximum likelihood estimation. Notably, the standard errors of the estimated parameters using FIL are consistently smaller, except for the IL method, which is affected by separation. As expected, the NRI method estimates the smallest value for $E_0$, resulting in the smallest maximum achievable effect ($E_{max} + E_0$) among all methods.

\begin{figure}[htbp]
    \centering
    \includegraphics[width=1\linewidth]{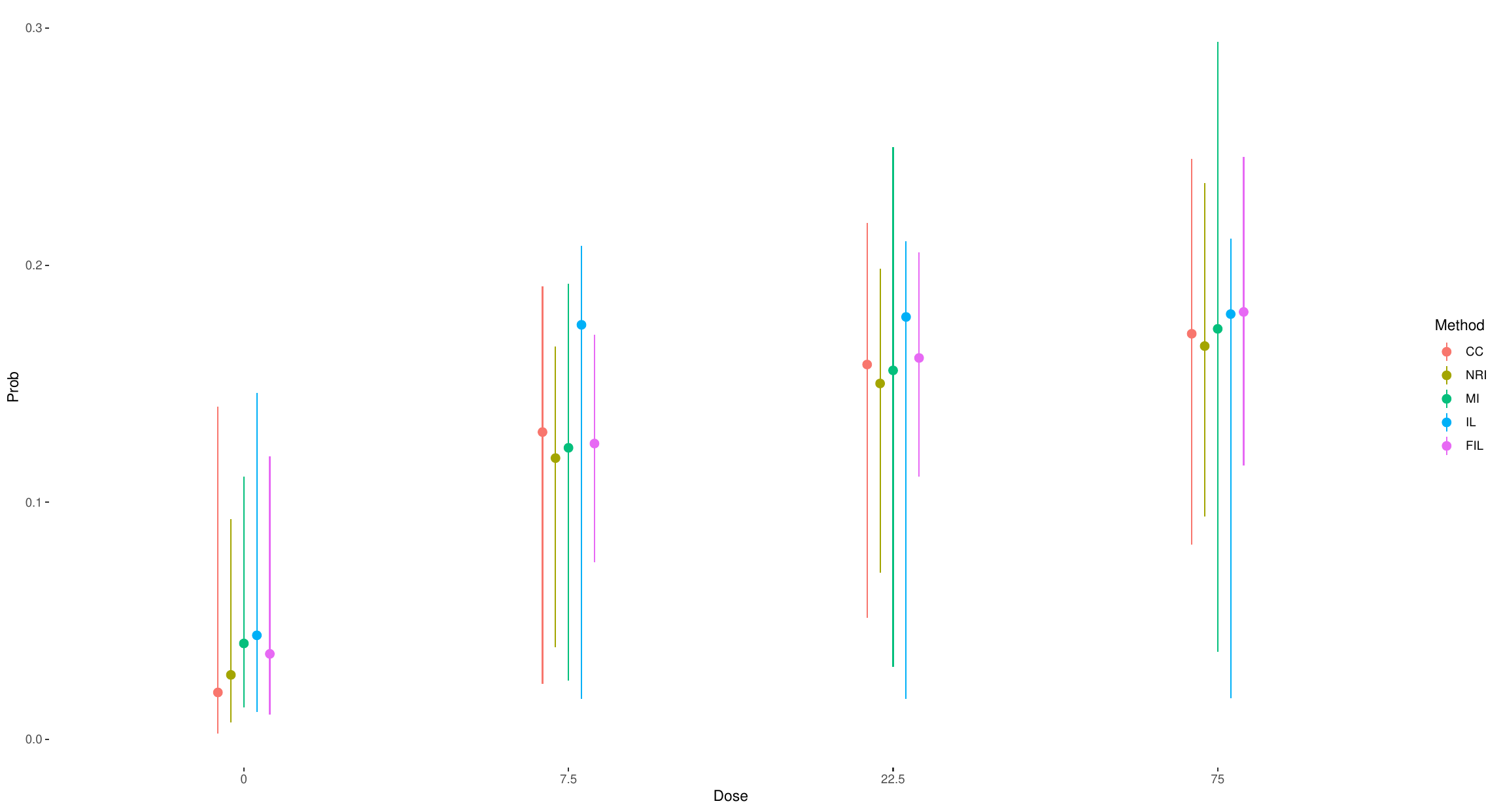}
    \caption{Estimated dose response remission probabilities based on different methods with their bootstrapped 95\% confidence intervals.}
    \label{fig:dose_response}
\end{figure}

To further evaluate the performance of these methods, we estimated the probabilities of remission and their bootstrapped 95\% confidence intervals using 5000 bootstrap samples, as shown in Figure \ref{fig:dose_response}. We observe that NRI consistently provides the lowest estimated probabilities across all dose groups due to its imputation strategy, but with relatively narrow confidence intervals. FIL, on the other hand, offers consistent estimations across all dose groups with the smallest variances. Due to the unstable estimation of $ED_{50}$, the IL method estimates the probability that nearly reaches the maximum effect at 7.5 mg, accompanied by comparably large and asymmetric confidence intervals across all dose groups. It is also noteworthy that the MI method exhibits the largest variance in estimations, despite its mean estimates being similar to those of FIL. This may be attributed to a violation of the missing mechanism assumption.

\textbf{Figure \ref{fig:dose_response} is approximately here.}

Table \ref{Table::missing_p_value} presents the parameter estimates from the logistic regression of the missingness indicator, including their standard errors, Z-values, and p-values. The response variable clinical remission, coded as y, is significant at the 5\% significance level while fitting the model in both the Ibrahim-Lipsitz (IL) and Firth-adjusted IL (FIL) methods, indicating that the probability of a response being missing depends on the response itself. This result supports the assumption that the missing data mechanism in this dataset is nonignorable.

\textbf{Table \ref{Table::missing_p_value} is approximately here.}

\section{Conclusion and Discussion}
In this article, we addressed the challenge of estimating the coefficients for the binary Emax model in the presence of missing responses under a nonignorable missing data mechanism. Our simulation studies demonstrate that the proposed Firth-type corrected weighted EM procedure of Ibrahim and Lipsitz (FIL) outperforms commonly used missing data handling strategies such as non-responder imputation (NRI) and multiple imputation (MI). Additionally, when fitting the binary Emax model with small or medium sample sizes, maximum likelihood-based approaches often face convergence issues due to complete separation or produce unstable estimates with significant bias and variation due to quasi-separation. In such scenarios, the FIL method offers a robust and reliable solution, effectively addressing the complexities and challenges posed by the data.

For both the IL and FIL methods, it is crucial to select an appropriate model for describing the likelihood of the missingness indicator, \( f(r_i \mid z_i, \ualpha) \), to accurately capture the underlying missing data mechanism. Standard model selection techniques, such as backward selection using AIC or likelihood ratio tests (as discussed by Ibrahim and Lipsitz (\cite{ibrahim1996parameter})), should be used in conjunction with considerations of scientific or clinical relevance. These techniques should be applied with the joint likelihood \( f(\utheta, \ualpha \mid z_i, r_i) \) while keeping \( f(y_i \mid Dose_{i}, \utheta\) fixed. Additionally, commonly used penalized likelihood variable selection methods, such as Lasso and Elastic-net regression, could be considered. In cases where the missingness mechanism is well separated by regression models, a Firth-corrected regression model, as implemented in the FIL method, may be appropriate. Our simulations indicate that the estimation of \( \ualpha \) can significantly impact the estimation of \( \utheta \) under the Emax model, particularly when separation occurs in predicting the missingness indicator. Therefore, it is advisable to examine the estimation results of \( \ualpha \) and consider adjustments if instability is observed.

The rationale for choosing an EM approach rather than multiple imputation (MI) to address missingness may warrant further explanation. As discussed in the introduction, MI is primarily designed for the missing at random (MAR) mechanism, which does not apply when the missing data are nonignorable. While recent literature, such as Im \& Kim (\cite{Im2017}) and Galimard et al. (\cite{Galimard2018}), has proposed methods to handle nonignorable missingness, the selection of an appropriate imputation model remains a challenge and is often determined through sensitivity analysis. In contrast, with the IL and FIL methods, model selection can be performed using standard techniques, making these methods more straightforward to implement and interpret.

In dose-response analysis, the Multiple Comparison Procedure-Modeling (MCP-Mod) approach, developed by Bretz, Pinheiro, and Branson (\cite{bretz2005combining,mcp-mod}), combines hypothesis testing and modeling with Type I error control. MCP-Mod uses AIC to select the best model for fitting the dose-response relationship, making it a natural extension to apply our proposed methods within the MCP-Mod framework, given its likelihood-based foundation. Indeed, Diniz et al. have recently developed a Firth-type MCP-Mod for Weibull regression with time-to-event data in small sample sizes (\cite{Diniz2023}). Further research could explore the implementation of these methods for binary or count dose-response models. In fact, MCP-mod contains a four-parameter Emax model as potential, thus extending to MCP-mod can capture situations when the three-parameter Emax model is misspecified. Additionally, penalization methods could be applied to the potential models within MCP-Mod to control the risk of separation, offering another avenue for future investigation.

\printbibliography

\newpage

\begin{table}[htbp]
\centering
{\scriptsize
 \resizebox*{!}{0.9\textheight}{
\begin{tabular}{cllcccccc}
  \arrayrulecolor{black}\hline\arrayrulecolor[gray]{0.8}
 Sample Size(N)&  Parameter &Type  & Estimate & MBE & MSE & Est.SE & CP & Est.Length\\ 
  \arrayrulecolor{black}\hline\arrayrulecolor[gray]{0.8}
 150 &log($ED_{50}$)& CC  & 2.104 & 0.089 & 0.461 & 0.698 & 0.979 & 2.736 \\ 
   & &NRI & 2.564 & 0.549 & 0.736 & 0.666 & 0.896 & 2.610 \\ 
   && MI  & 2.203 & 0.189 & 0.437 & 1.054 & 0.954 & 4.132 \\ 
   & &IL  & 2.026 & 0.011 & 0.581 & 0.738 & 0.982 & 2.895 \\ 
   & &FIL  & 2.099 & 0.084 & 0.338 & 0.597 & 0.973 & 2.341 \\ 
  \hline
  &$E_{max}$& CC  & 3.900 & 0.316 & 0.551 & 0.862 & 0.989 & 3.377 \\ 
   & &NRI  & 4.185 & 0.602 & 0.883 & 0.877 & 0.974 & 3.438 \\ 
   && MI  & 3.841 & 0.258 & 0.542 & 0.803 & 0.974 & 3.147 \\ 
   & &IL  & 3.704 & 0.121 & 0.567 & 0.851 & 0.964 & 3.338 \\ 
   & &FIL  & 3.626 & 0.043 & 0.365 & 0.766 & 0.976 & 3.004 \\ 
    \hline
  &$E_0$& CC  & -2.396 & -0.199 & 0.395 & 0.781 & 0.973 & 3.061 \\ 
   & &NRI  & -2.618 & -0.421 & 0.513 & 0.740 & 0.991 & 2.901 \\
   && MI  & -2.339 & -0.141 & 0.344 & 0.663 & 0.966 & 2.600 \\ 
   & &IL  & -2.199 & -0.002 & 0.413 & 0.749 & 0.928 & 2.936 \\ 
   & &FIL  & -2.177 & 0.002 & 0.263 & 0.686 & 0.954 & 2.691 \\ 
   \arrayrulecolor{black}\hline\arrayrulecolor[gray]{0.8}
  
250 &log($ED_{50}$)& CC  & 2.058 & 0.043 & 0.279 & 0.521 & 0.968 & 2.042 \\ 
   && NRI  & 2.486 & 0.471 & 0.488 & 0.501 & 0.849 & 1.964 \\ 
   && MI  & 2.181 & 0.166 & 0.276 & 0.473 & 0.942 & 1.853 \\ 
   && IL  & 1.964 & -0.051 & 0.597 & 0.535 & 0.970 & 2.097 \\ 
   && FIL  & 2.032 & 0.017 & 0.222 & 0.479 & 0.966 & 1.878 \\ 
   \hline
&$E_{max}$& CC  & 3.898 & 0.315 & 0.500 & 0.667 & 0.979 & 2.614 \\ 
   && NRI  & 4.137 & 0.554 & 0.639 & 0.627 & 0.947 & 2.456 \\
   && MI  & 3.858 & 0.275 & 0.478 & 0.587 & 0.948 & 2.301 \\ 
   && IL  & 3.740 & 0.156 & 0.446 & 0.660 & 0.967 & 2.587 \\ 
   & &FIL & 3.678 & 0.095 & 0.337 & 0.619 & 0.971 & 2.428 \\ 
    \hline
  & $E_0$ & CC  & -2.451 & -0.254 & 0.427 & 0.623 & 0.980 & 2.443 \\ 
   & &NRI  & -2.668 & -0.470 & 0.546 & 0.591 & 0.996 & 2.319 \\ 
   && MI  & -2.386 & -0.189 & 0.375 & 0.530 & 0.958 & 2.078 \\ 
   & &IL  & -2.303 & -0.106 & 0.397 & 0.614 & 0.955 & 2.406 \\ 
   & &FIL  & -2.260 & -0.062 & 0.282 & 0.569 & 0.966 & 2.232 \\  
   \arrayrulecolor{black}\hline\arrayrulecolor[gray]{0.8}

  350 & log($ED_{50}$)&  CC & 2.087 & 0.072 & 0.201 & 0.436 & 0.959 & 1.708 \\ 
   & &NRI& 2.514 & 0.499 & 0.454 & 0.423 & 0.783 & 1.657 \\ 
   & & MI  & 2.162 & 0.147 & 0.200 & 0.394 & 0.925 & 1.543 \\
   & &IL  & 2.005 & -0.010 & 0.211 & 0.449 & 0.967 & 1.759 \\ 
   & &FIL  & 2.041 & 0.026 & 0.170 & 0.414 & 0.962 & 1.624 \\  
   \hline
  &$E_{max}$& CC  & 3.863 & 0.280 & 0.422 & 0.553 & 0.962 & 2.169 \\ 
   & &NRI  & 4.106 & 0.523 & 0.553 & 0.519 & 0.896 & 2.034 \\ 
   & & MI  & 3.816 & 0.232 & 0.364 & 0.485 & 0.937 & 1.901 \\ 
   & &IL  & 3.712 & 0.128 & 0.393 & 0.550 & 0.944 & 2.155 \\ 
   & &FIL  & 3.670 & 0.086 & 0.315 & 0.526 & 0.949 & 2.060 \\ 
   \hline
  &$E_0$& CC & -2.436 & -0.238 & 0.364 & 0.520 & 0.984 & 2.037 \\ 
   & &NRI  & -2.653 & -0.455 & 0.485 & 0.493 & 0.958 & 1.933 \\ 
   & &MI  & -2.365 & -0.168 & 0.297 & 0.440 & 0.955 & 1.725 \\ 
   & &IL  & -2.295 & -0.098 & 0.341 & 0.514 & 0.942 & 2.017 \\ 
   & &FIL  & -2.265 & -0.067 & 0.260 & 0.487 & 0.957 & 1.909 \\ 
   \arrayrulecolor{black}\hline\arrayrulecolor[gray]{0.8}

 450 & log($ED_{50}$)& CC  & 2.077 & 0.062 & 0.158 & 0.378 & 0.947 & 1.482 \\ 
   && NRI  & 2.499 & 0.484 & 0.395 & 0.367 & 0.750 & 1.438 \\
   && MI  & 2.156 & 0.141 & 0.158 & 0.340 & 0.916 & 1.333\\
   && IL  & 1.998 & -0.017 & 0.161 & 0.389 & 0.950 & 1.524 \\ 
   && FIL  & 2.025 & 0.010 & 0.137 & 0.366 & 0.953 & 1.435 \\ 
   \hline
  &$E_{max}$& CC  & 3.863 & 0.280 & 0.320 & 0.481 & 0.962 & 1.884 \\ 
   && NRI  & 4.099 & 0.516 & 0.463 & 0.448 & 0.869 & 1.757 \\
   && MI &  3.794 & 0.210 & 0.279 & 0.421 & 0.916 & 1.650 \\
   && IL  & 3.712 & 0.128 & 0.280 & 0.478 & 0.961 & 1.875 \\ 
   && FIL  & 3.682 & 0.098 & 0.233 & 0.463 & 0.966 & 1.813 \\ 
  \hline
  &$E_0$& CC  & -2.428 & -0.231 & 0.284 & 0.452 & 0.974 & 1.772 \\ 
   && NRI  & -2.644 & -0.446 & 0.403 & 0.428 & 0.925 & 1.680 \\ 
   && MI  & -2.358 & -0.161 & 0.242 & 0.381 & 0.932 & 1.495\\
   && IL  & -2.287 & -0.090 & 0.255 & 0.448 & 0.953 & 1.756 \\ 
   && FIL  & -2.266 & -0.068 & 0.206 & 0.430 & 0.956 & 1.685 \\ 
   \arrayrulecolor{black}\hline\arrayrulecolor[gray]{0.8}
 
\end{tabular}
}
}
\caption{Estimates, mean bias error, mean squared error, estimated standard errors, coverage probabilities, and 95\% Wald confidence intervals based on 1000 simulations with missing rate$\approx$15\%.}
\label{Table::diff_N}
\end{table}

\begin{table}[htbp]
    \centering
    \begin{tabular}{lccccc}
    \hline
        Dose(mg) & Placebo & 7.5 & 22.5 & 75 & 225 \\
        \hline
         Sample size& 73 & 71 & 72 & 71 & 70\\
         Missing response & 6 &8 &1 & 3& 6\\
         Remission (Yes) & 2 & 8& 12 & 11 & 4\\
         Previous TNF therapy (Yes) & 42 & 41 & 41 & 41 & 40\\
         Sex (Male) & 44 &39 &46 &37&42\\
         IS (Yes) & 15 & 23 & 23 &21 &20\\
         SD (Yes) & 31 & 38 &38 &36 &36 \\
         ASA (Yes) & 47 & 37&35 &45 &35\\
        MCSBASE (Mean) & 8.425 & 8.732 & 8.083 & 8.380 & 8.686\\
         CRPBASE (Mean) & 1.106 & 1.097 & 1.149 & 0.979 & 0.892\\
         Age (Mean) & 38.616 & 41.310 & 42.222 & 37.465 & 41.300\\
         \hline
    \end{tabular}
    \caption{Sample size, the number of missing response cases, the number of remission cases, and baseline statistics for each dosage group in TURANDOT study.}
    \label{Table::turandot_descriptive}
\end{table}

\begin{table}[htp]
    \centering
    {\scriptsize
    \begin{tabular}{lcccc}
    \hline
         Parameter& Method& Estimate & StdErr & 95\% CI  \\
        \arrayrulecolor{black}\hline\arrayrulecolor[gray]{0.8}
         log($ED_{50}$)& CC & 0.480 & 1.856 & (-3.159, 4.119)\\
                     & NRI & 0.756 & 1.484 & (-2.153, 3.664)\\
                     & MI & 0.462 & 4.611 & (-8.576, 9.500)\\
                     & IL & -1.775 & 15.065 & (-31.303, 27.752)\\
                     & FIL & 1.030 & 0.907 & (-0.747, 2.808)\\
        \hline
         $E_{max}$& CC & 1.938 & 0.788 & (0.394, 3.481)\\
                & NRI & 2.017 & 0.788 & (0.472, 3.561)\\
                & MI & 1.851 & 0.741 & (0.400,3.303)\\
                & IL & 1.563 & 0.748 & (0.098, 3.029)\\
                & FIL & 1.836 & 0.721 & (0.423, 3.249)\\
        \hline
         $E_{0}$& CC & -3.484 & 0.718 & (-4.890, -2.077)\\
              & NRI & -3.576 & 0.716 & (-4.980, -2.172)\\
              & MI & -3.400 & 0.667 & (-4.707, -2.092)\\
              & IL & -3.080 & 0.685 & (-4.423, -1.738)\\
              & FIL & -3.285 & 0.642 & (-4.543, -2.027)\\
       \arrayrulecolor{black}\hline\arrayrulecolor[gray]{0.8}
    \end{tabular}
    }
     \caption{Analysis result of TURANDOT data with different missing data handling methods and the proposed method.}
     
    \label{Table::turandot_analysis}
\end{table}

\begin{table}[htbp]
    \centering
    {\scriptsize
    \begin{tabular}{cccccc}
    \arrayrulecolor{black}\hline\arrayrulecolor[gray]{0.8}
        Method & Variable &Estimate &StdErr & Z-value & p-value  \\
        \arrayrulecolor{black}\hline\arrayrulecolor[gray]{0.8}
         IL & intercept &-4.006 &2.233 &-1.794 &0.073\\
          & y &2.532 & 0.595 & 4.254 & $<${\textbf 0.001}\\
          & Dose &-0.027 & 0.016 & -1.726 & 0.084\\
          &MCSBASE &0.222 & 0.202 &1.098 & 0.272\\
          &CRPBASE &0.260 & 0.127 & 2.042 & 0.041\\
          & AGE &-0.042 & 0.027 & -1.550 & 0.121\\
          & Sex (Male) &1.083 & 0.680 & 1.592 & 0.113\\
          & IS (Yes) &-0.855 & 0.840 & -1.018 & 0.309\\
          & SD (Yes) &0.481 & 0.671 & -0.915 & 0.360\\
          & ASA (Yes)&-0.614 & 0.676 &0.710 & 0.478\\
          \hline
        FIL & intercept &-2.510 &1.824 &-1.375 & 0.169\\
          & y & 1.063 & 0.493 &2.156 & {\textbf 0.031}\\
          & Dose & -0.015 & 0.011 &-1.396 & 0.163\\
          &MCSBASE &0.123 & 0.169 & 0.726 &0.467\\
          &CRPBASE &0.223 & 0.113 & 1.975 & 0.048\\
          & AGE & -0.038 & 0.022 & -1.735 & 0.083\\
          & Sex (Male) &0.967 & 0.551 & 1.756 & 0.079\\
          & IS (Yes) & -0.663 & 0.657 & -1.008 &0.313\\
          & SD (Yes) &0.250 & 0.550 & 0.455 & 0.650\\
          & ASA (Yes) & -0.642 & 0.550 & -1.168 & 0.243\\
          \arrayrulecolor{black}\hline\arrayrulecolor[gray]{0.8}
    \end{tabular}
    }
    \caption{Estimates of missing data model for TURANDOT study. }
    \label{Table::missing_p_value}
\end{table}

\end{document}

\maketitle

\begin{table}[htbp]
\centering
{\scriptsize
\resizebox*{!}{0.9\textheight}{
\begin{tabular}{cllcccccc}
  \arrayrulecolor{black}\hline\arrayrulecolor[gray]{0.8}
 Sample Size(N)&  Parameter &Type  & Estimate & MBE & MSE & Est.SE & CP & Est.Length\\ 
  \arrayrulecolor{black}\hline\arrayrulecolor[gray]{0.8}
   250 &log($ED_{50}$)& CC &   1.900 & -0.115 & 0.212 & 0.568 & 0.990 & 2.225 \\ 
   && NRI &   1.548 & -0.467 & 0.353 & 0.430 & 0.890 & 1.684 \\ 
   && MI & 2.271 & 0.256 & 0.433 & 0.486 & 0.967 & 1.905 \\
   && IL &   2.998 & 0.983 & 4.480 & 0.703 & 0.687 & 2.758 \\ 
   && FIL &   2.219 & 0.204 & 0.463 & 0.568 & 0.960 & 2.227 \\ 
   \hline
   
   &$E_{max}$& CC & 4.057 & 0.473 & 0.494 & 0.915 & 1.000 & 3.585 \\ 
   && NRI & 5.105 & 1.522 & 2.536 & 0.928 & 0.813 & 3.636 \\
   && MI  & 3.632 & 0.048 & 0.288 & 0.581 & 0.967 & 2.277 \\
   && IL & 3.853 & 0.270 & 1.589 & 1.043 & 0.807 & 4.087 \\ 
   && FIL & 3.594 & 0.011 & 0.362 & 0.782 & 0.957 & 3.065 \\ 
   \hline
    
   &$E_0$& CC &   -2.623 & 0.532 & 0.378 & 0.925 & 1.000 & 3.628 \\ 
   && NRI    & -3.706 & -1.508 & 2.433 & 0.920 & 0.997 & 3.606 \\ 
   && MI &  -2.104 & 0.124 & 0.199 & 0.611 & 0.930 & 2.004 \\ 
   && IL & -1.921 & 0.276 & 2.159 & 0.703 & 0.657 & 2.613 \\ 
   && FIL & -2.117 & 0.080 & 0.372 & 0.568 & 0.957 & 2.920 \\ 
   \arrayrulecolor{black}\hline\arrayrulecolor[gray]{0.8}

  350 & log($ED_{50}$)& CC   & 1.783 & -0.198 & 0.197& 0.476& 0.970& 1.865\\ 
  & & NRI   & 1.476 & -0.523 & 0.382 & 0.367& 0.742& 1.438\\ 
  && MI  & 2.048 & 0.233 & 0.126 & 0.493 & 0.990 & 1.541 \\
  & & IL   & 2.822 & 0.762 & 4.074& 0.590& 0.840& 2.312\\ 
  & & FIL   & 2.092 & 0.116 & 0.275& 0.497& 0.952& 1.949\\ 
   \hline
  
   &$E_{max}$ & CC   & 4.283 & 0.626 & 0.699& 0.875& 0.998& 3.432\\ 
   && NRI  & 5.333 & 1.690 & 3.129 & 0.887& 0.498& 3.478\\ 
   && MI  & 3.833 & 0.249 & 0.347 & 0.522 & 0.950 & 2.047 \\ 
   && IL &   3.859  & 0.200 & 1.876 & 0.912& 0.684& 3.575\\ 
   && FIL   & 3.681 & 0.132 & 0.447 & 0.727& 0.954& 2.848\\ 
    \hline
    
  & $E_0$ & CC  & -2.885 & -0.607 & 0.632& 0.888& 0.998& 3.480\\ 
   && NRI  & -3.957 & -1.697 & 3.098& 0.885& 0.330& 3.469\\ 
   && MI  & -2.395 & -0.197 & 0.274 & 0.480 & 0.943 & 1.881 \\ 
   && IL   & -2.054 & 0.174 & 2.141 & 0.646& 0.484& 2.534\\ 
   && FIL   & -2.250 & -0.035 & 0.414& 0.722& 0.942& 2.830\\ 
   \arrayrulecolor{black}\hline\arrayrulecolor[gray]{0.8}
  
   450 &log($ED_{50}$)& CC   & 1.786 & -0.229 & 0.187 & 0.419 & 0.960 & 1.644 \\ 
   && NRI &   1.477 & -0.538 & 0.368 & 0.325 & 0.663 & 1.273 \\
    && MI  & 2.143 & 0.128 & 0.120 & 0.341 & 0.960 & 1.339 \\
   && IL &   2.564 & 0.549 & 2.580 & 0.508 & 0.553 & 1.991 \\ 
   && FIL &   2.168 & 0.153 & 0.357 & 0.457 & 0.943 & 1.793 \\ 
   \hline
   
   &$E_{max}$& CC   & 4.286 & 0.702 & 0.806 & 0.783 & 1.000 & 3.071 \\ 
   && NRI   & 5.332 & 3.135 & 1.749 & 3.371 & 0.230 & 3.111 \\
   && MI  & 3.867 & 0.284 & 0.352 & 0.462 & 0.913 & 1.810 \\
   && IL   & 3.880 & 0.297 & 1.747 & 0.758 & 0.573 & 2.972 \\ 
   && FIL   & 3.627 & 0.043 & 0.450 & 0.639 & 0.953 & 2.503 \\ 

   \hline
   &$E_0$& CC   & -2.885 & -0.688 & 0.774 & 0.793 & 0.997 & 3.108 \\ 
   && NRI   & -3.953 & -1.756 & 3.369 & 0.792 & 0.127 & 3.103 \\
   && MI  & -2.419 & -0.222 & 0.295 & 0.425 & 0.933 & 1.664 \\
   && IL &   -2.196 & 0.021 & 2.006  & 0.595 & 0.727 & 2.333 \\ 
   && FIL   & -2.176 & 0.022 & 0.502 & 0.628 & 0.956 & 2.461 \\ 
   \arrayrulecolor{black}\hline\arrayrulecolor[gray]{0.8}
\end{tabular}
}
}
\caption{Estimates, absolute bias, mean squared error, estimated standard errors, coverage probabilities and 95\% Wald confidence intervals based on 1000 simulations with missing rate$\approx$15\% with uneven missing pattern.}
\label{Table::S1}
\end{table}

\begin{table}[htbp]
\centering
{\scriptsize
\resizebox*{!}{0.9\textheight}{
\begin{tabular}{cllcccccc}
  \arrayrulecolor{black}\hline\arrayrulecolor[gray]{0.8}
 Missing rate (\%)&  Parameter &Type  & Estimate & MBE & MSE & Est.SE & CP & Est.Length\\ 
  \arrayrulecolor{black}\hline\arrayrulecolor[gray]{0.8}
  10 & log($ED_{50}$)&CC  & 2.014 & -0.001 & 0.182 & 0.418 & 0.956 & 1.637 \\ 
  & & NRI  & 2.275 & 0.260 & 0.260 & 0.413 & 0.898 & 1.619 \\ 
  & &MI & 2.076 & 0.061 & 0.179 & 0.394 & 0.944 & 1.543 \\ 
  & & IL  & 1.979 & -0.036 & 0.192 & 0.422 & 0.955 & 1.656 \\ 
  & & FIL  & 2.009 & -0.006 & 0.156 & 0.393 & 0.956 & 1.540 \\ 
   \hline
  &$E_{max}$& CC  & 3.742 & 0.158 & 0.316 & 0.509 & 0.963 & 1.995 \\ 
  & & NRI  & 3.833 & 0.249 & 0.321 & 0.489 & 0.952 & 1.918 \\ 
  & & MI  & 3.699 & 0.116 & 0.264 & 0.467 & 0.948 & 1.829 \\
  & & IL  & 3.678 & 0.094 & 0.304 & 0.506 & 0.958 & 1.984 \\ 
  & & FIL  & 3.639 & 0.056 & 0.255 & 0.489 & 0.961 & 1.916 \\ 
   \hline
  &$E_0$& CC  & -2.346 & -0.149 & 0.282 & 0.469 & 0.976 & 1.837 \\ 
  & & NRI  & -2.482 & -0.285 & 0.328 & 0.458 & 0.974 & 1.794 \\ 
  & &MI &  -2.277 & -0.080 & 0.217 & 0.419 & 0.950 & 1.641 \\ 
  & & IL  & -2.282 & -0.085 & 0.269 & 0.465 & 0.957 & 1.823 \\ 
  & & FIL  & -2.255 & -0.058 & 0.216 & 0.445 & 0.964 & 1.746 \\ 
   \arrayrulecolor{black}\hline\arrayrulecolor[gray]{0.8}
  
15 & log($ED_{50}$)&  CC & 2.087 & 0.072 & 0.201 & 0.436 & 0.959 & 1.708 \\ 
   & &NRI& 2.514 & 0.499 & 0.454 & 0.423 & 0.783 & 1.657 \\ 
   & & MI  & 2.162 & 0.147 & 0.200 & 0.394 & 0.925 & 1.543 \\
   & &IL  & 2.005 & -0.010 & 0.211 & 0.449 & 0.967 & 1.759 \\ 
   & &FIL  & 2.041 & 0.026 & 0.170 & 0.414 & 0.962 & 1.624 \\  
   \hline
  &$E_{max}$& CC  & 3.863 & 0.280 & 0.422 & 0.553 & 0.962 & 2.169 \\ 
   & &NRI  & 4.106 & 0.523 & 0.553 & 0.519 & 0.896 & 2.034 \\ 
   & & MI  & 3.816 & 0.232 & 0.364 & 0.485 & 0.937 & 1.901 \\ 
   & &IL  & 3.712 & 0.128 & 0.393 & 0.550 & 0.944 & 2.155 \\ 
   & &FIL  & 3.670 & 0.086 & 0.315 & 0.526 & 0.949 & 2.060 \\ 
   \hline
  &$E_0$& CC & -2.436 & -0.238 & 0.364 & 0.520 & 0.984 & 2.037 \\ 
   & &NRI  & -2.653 & -0.455 & 0.485 & 0.493 & 0.958 & 1.933 \\ 
   & &MI  & -2.365 & -0.168 & 0.297 & 0.440 & 0.955 & 1.725 \\ 
   & &IL  & -2.295 & -0.098 & 0.341 & 0.514 & 0.942 & 2.017 \\ 
   & &FIL  & -2.265 & -0.067 & 0.260 & 0.487 & 0.957 & 1.909 \\ 
   \arrayrulecolor{black}\hline\arrayrulecolor[gray]{0.8}

25 & log($ED_{50}$)& CC  &2.215 & 0.200 & 0.278 & 0.477 & 0.939 & 1.871 \\ 
  & & NRI  &2.945 & 0.930 & 1.173 & 0.451 & 0.480 & 1.770 \\ 
  && MI & 2.288 & 0.273 & 0.259 & 0.384 & 0.872 & 1.507 \\
  & & IL  &2.076 & 0.061 & 0.836 & 0.523 & 0.956 & 2.049 \\ 
  & & FIL  &2.064 & 0.049 & 0.226 & 0.465 & 0.950 & 1.821 \\ 
  \hline
  &$E_{max}$&  CC  & 4.164 & 0.581 & 0.760 & 0.678 & 0.961 & 2.656 \\ 
  & & NRI  &4.696 & 1.112 & 1.572 & 0.599 & 0.546 & 2.348 \\
  & &MI  & 4.161 & 0.578 & 0.693 & 0.544 & 0.862 & 2.131 \\ 
  & & IL  &3.719 & 0.135 & 0.672 & 0.695 & 0.918 & 2.725 \\ 
  & & FIL  &3.674 & 0.091 & 0.410 & 0.620 & 0.945 & 2.431 \\ 
   \hline
  &$E_0$& CC  &-2.706 & -0.509 & 0.657 & 0.662 & 0.993 & 2.596 \\ 
  & & NRI  &-3.060 & -0.863 & 1.047 & 0.563 & 0.790 & 2.205 \\ 
  & &MI  & -2.658 & -0.460 & 0.550 & 0.507 & 0.940 & 1.988 \\
  & & IL  &-2.247 & -0.050 & 0.585 & 0.635 & 0.893 & 2.487 \\ 
  & & FIL  &-2.262 & -0.065 & 0.355 & 0.592 & 0.941 & 2.321 \\ 
  \arrayrulecolor{black}\hline\arrayrulecolor[gray]{0.8}
  
  30 & log($ED_{50}$)& CC  &2.244 & 0.229 & 0.318 & 0.505 & 0.921 & 1.980 \\ 
  & & NRI  & 3.339 & 1.324 & 2.181 & 0.569 & 0.365 & 2.231 \\ 
  & &MI  & 2.392 & 0.377 & 0.373 & 0.418 & 0.830 & 1.640 \\ 
  & & IL  &2.099 & 0.085 & 1.149 & 0.533 & 0.961 & 2.088 \\ 
  & & FIL  &2.038 & 0.023 & 0.227 & 0.475 & 0.958 & 1.860 \\ 
   \hline
   &$E_{max}$& CC  & 4.123 & 0.540 & 0.656 & 0.681 & 0.979 & 2.670 \\ 
  & & NRI  & 4.438 & 0.854 & 1.080 & 0.640 & 0.812 & 2.507 \\ 
   && MI  & 4.032 & 0.449 & 0.607 & 0.545 & 0.892 & 2.136 \\ 
  & & IL  & 3.740 & 0.156 & 0.640 & 0.785 & 0.937 & 3.077 \\ 
  & & FIL  & 3.673 & 0.089 & 0.361 & 0.626 & 0.956 & 2.455 \\ 
  \hline
  &$E_0$& CC  & -2.705 & -0.508 & 0.626 & 0.665 & 0.996 & 2.607 \\ 
  & & NRI  & -2.843 & -0.646 & 0.683 & 0.528 & 0.921 & 2.072 \\ 
  & & MI  & -2.604 & -0.406 & 0.513 & 0.499 & 0.950 & 1.955 \\ 
  & & IL  & -2.263 & -0.066 & 0.577 & 0.635 & 0.899 & 2.490 \\ 
  & & FIL  & -2.230 & -0.033 & 0.333 & 0.597 & 0.942 & 2.340 \\  
   \arrayrulecolor{black}\hline\arrayrulecolor[gray]{0.8}
 
\end{tabular}
}
}
\caption{Estimates, mean bias error, mean squared error, estimated standard errors, coverage probabilities, and 95\% Wald confidence intervals based on 1000 simulations with sample size n=350.}
\label{Table::S2}
\end{table}